\documentclass{article}

\usepackage{epsfig}
\usepackage{amsmath,theorem}
\usepackage{xspace}
\usepackage{amssymb}
\usepackage{euscript}
\usepackage{named}

\newcommand{\KB}{KB\xspace}

\newcommand{\DL}{\textrm{DL}\xspace}

\newfont{\mymathtt}{cmtt10 scaled 1095}

\newcommand{\scname}[1]{\textsc{#1}}

\newcommand{\galen}{\scname{Galen}\xspace}
\newcommand{\Galen}{\galen}

\newcommand{\alc}{
  \ensuremath{\mathcal{ALC}}\xspace}

\newcommand{\alcRtrans}{
        \ensuremath{\alc_{R^{+}}}\xspace}
\newcommand{\s}{
        \ensuremath{\mathcal{S}}\xspace}

\newcommand{\si}{
  \ensuremath{\mathcal{SI}}\xspace}

\newcommand{\shi}{
  \ensuremath{\mathcal{SHI}}\xspace}

\newcommand{\sh}{
  \ensuremath{\mathcal{SH}}\xspace}

\newcommand{\shif}{
  \ensuremath{\mathcal{SHIF}}\xspace}              

\renewcommand{\sin}{
  \ensuremath{\mathcal{SIN}}\xspace}

\newcommand{\shin}{
  \ensuremath{\mathcal{SHIN}}\xspace} 
\newcommand{\shiq}{
  \ensuremath{\mathcal{SHIQ}}\xspace} 

\newcommand{\Fact}{FaCT\xspace}
\newcommand{\iFaCT}{iFaCT\xspace}
\newcommand{\IFact}{\iFaCT}
\newcommand{\Dlp}{DLP\xspace}
\newcommand{\PDL}
        {PDL\xspace}
\newcommand{\CPDL}
        {\emph{converse}-PDL\xspace}

\newfont{\bigmathxx}{cmsy10 scaled 1440}
\newfont{\smallmathxx}{cmsy10 scaled 720}

\newcommand{\bigsqcap}{\mathop{\mathop{\mbox{\bigmathxx\symbol{117}}}}\limits}

\newcommand{\I}{
        \ensuremath{\mathcal{I}}\xspace}

\newcommand{\Tree}{
        \ensuremath{\mathbf{T}}\xspace}
\newcommand{\ifunc}{
        \ensuremath{^\mathcal{I}}\xspace}
\newcommand{\deltai}{
        \ensuremath{\Delta\ifunc}\xspace}

\newcommand{\Pspace}{\textsc{Pspace}\xspace}

\newcommand{\Term}{
        \ensuremath{\mathcal{T}}\xspace}

\newcommand{\Roles}{\ensuremath{\mathbf{R}}\xspace}
\newcommand{\Rplus}{\ensuremath{\mathbf{R}_+}\xspace}

\newcommand{\Lab}{\ensuremath{\EuScript{L}}\xspace}
\newcommand{\BLab}{\ensuremath{\EuScript{B}}\xspace}
\newcommand{\Edges}{\ensuremath{\EuScript{E}}\xspace}
\newcommand{\Inv}{\mathop{\mathsf{Inv}}}
\newcommand{\Tr}{\mathop{\mathsf{Trans}}}

\newcommand{\K}
        {\ensuremath{\mathbf{K}}\xspace}
\newcommand{\KT}
        {\ensuremath{\mathbf{KT}}\xspace}
\newcommand{\Kfour}
        {\ensuremath{\mathbf{K4}}\xspace}
\newcommand{\Sfour}
        {\ensuremath{\mathbf{S4}}\xspace}
\newcommand{\Km}
        {\ensuremath{\K_{(\mathbf{m})}}\xspace}
\newcommand{\KTm}
        {\ensuremath{\KT_{(\mathbf{m})}}\xspace}
\newcommand{\Kfourm}
        {\ensuremath{\mathbf{K4}_{(\mathbf{m})}}\xspace}
\newcommand{\Sfourm}
        {\ensuremath{\mathbf{S4}_{(\mathbf{m})}}\xspace}

\newcommand{\Not}{\neg}
\newcommand{\some}[2]{%
  \ensuremath{\exists #1 . #2}}
\newcommand{\Some}[2]{\some #1 #2}
\newcommand{\all}[2]{%
  \ensuremath{\forall #1 . #2}}
\newcommand{\All}[2]{\all #1 #2}
\newcommand{\Atmost}[2]{%
  \ensuremath{\mathopen{\leqslant} #1 #2}}
\newcommand{\Atleast}[2]{%
  \ensuremath{\mathopen{\geqslant} #1 #2}}
\newcommand{\AtmostQ}[3]{%
  \ensuremath{\mathopen{\leqslant} #1 #2.#3}}
\newcommand{\AtleastQ}[3]{%
  \ensuremath{\mathopen{\geqslant} #1 #2.#3}}
\newcommand{\atleast}[2]{%
  \ensuremath{\Atleast #1 #2}}
\newcommand{\atmost}[2]{%
  \ensuremath{\Atmost #1 #2}}
\newcommand{\atleastq}[3]{%
  \ensuremath{\AtleastQ #1 #2 #3}}
\newcommand{\atmostq}[3]{%
  \ensuremath{\AtmostQ #1 #2 #3}}

\newcommand{\feat}[1]{\ensuremath{(\leqslant 1 \; #1)}}
\newcommand{\Negfeat}[1]{\ensuremath{(\geqslant 2 \; #1)}}

\newcommand{\tuple}[2]{
        \ensuremath{\langle #1 , #2 \rangle}}

\newcommand{\Card}[1]{
        \ensuremath{\vert#1\vert}}

\newcommand{\N}{\mathbb{N}}

\newcommand{\sss}{{\mathrel{\kern.25em{\sqsubseteq}\kern-.5em \mbox{{\scriptsize *}}\kern.25em}}}
\newcommand{\R}{\ensuremath{\mathcal{R}}\xspace}

\theoremstyle{break}
\theoremheaderfont{\bf\boldmath}
\newtheorem{satz}{Satz}[section]
\newtheorem{Theorem}[satz]{Theorem}
\newtheorem{Definition}[satz]{Definition}
\theoremstyle{plain}
\newtheorem{Lemma}[satz]{Lemma}

\usepackage{pstricks,pst-node}

\title{Practical Reasoning for Very Expressive Description Logics\thanks{This
    paper appeared in the Logic Journal of the IGPL 8(3):239-264, May 2000.}}

\author{
  {\bf Ian Horrocks}\\
  Department of Computer Science, University of Manchester\\
  \texttt{horrocks@cs.man.ac.uk} \and
  {\bf Ulrike Sattler}\\
  LuFG Theoretical Computer Science, RWTH Aachen\\
  \texttt{sattler@informatik.rwth-aachen.de} \and
  {\bf Stephan Tobies}\\
  LuFG Theoretical Computer Science, RWTH Aachen\\
  \texttt{tobies@informatik.rwth-aachen.de}
  }

\date{}

\begin{document}

\maketitle

\begin{abstract}
  Description Logics (DLs) are a family of knowledge representation formalisms
  mainly characterised by constructors to build complex concepts and roles from
  atomic ones. Expressive role constructors are
  important in many applications, but can be computationally
  problematical.
  
  We present an algorithm that decides satisfiability of the DL \alc extended
  with transitive and inverse roles and functional restrictions with respect to
  general concept inclusion axioms and role hierarchies; early experiments
  indicate that this algorithm is well-suited for implementation.  Additionally,
  we show that \alc extended with just transitive and inverse roles is still in
  \textsc{PSpace}.  We investigate the limits of decidability for this family of
  DLs, showing that relaxing the constraints placed on the kinds of roles used
  in number restrictions leads to the undecidability of all inference problems.
  Finally, we describe a number of optimisation techniques that are crucial in
  obtaining implementations of the decision procedures, which, despite the hight
  worst-case complexity of the problem, exhibit good performance with real-life
  problems.
\end{abstract}

\section{Motivation}\label{sec:motivat}

Description Logics (DLs) are a well-known family of knowledge representation
formalisms~\cite{DLNS96}. They are based on the notion of concepts (unary
predicates, classes) and roles (binary relations), and are mainly characterised
by constructors that allow complex concepts and roles to be built from atomic ones.  Sound and
complete algorithms for the interesting inference problems such as subsumption
and satisfiability of concepts are known for a wide variety of
DLs.

Transitive and inverse roles play an important role not only in the adequate
representation of complex, aggregated objects~\cite{HoSat98c}, but also for
reasoning with conceptual data models~\cite{CaLN94}. Moreover, defining concepts
using general concept inclusion axioms seems natural
and is crucial for representing conceptual data models.

The relevant inference problems for (an extension of) \alc augmented in the
described manner are known to be decidable~\cite{GiLe96}, and worst-case optimal
inference algorithms have been described~\cite{DeGiaMass98}. However, to the
best of our knowledge, nobody has found efficient means to deal with their high
degree of non-determinism, which so far prohibits their use in realistic
applications.  This is mainly due to the fact that these algorithms can handle
not only transitive roles but also the transitive closure of roles. It has been
shown~\cite{Sat95ashort} that restricting the DL to transitive roles can lead to
a lower complexity, and that transitive roles, even when combined with role
hierarchies, allow for algorithms that behave quite well in realistic
applications~\cite{Horrocks98c}.  However, until now it has been unclear if this
is still true when inverse roles are also present.

In this paper we present various aspects of our research in this direction.
Firstly, we motivate our use of logics with transitive roles instead of
transitive closure by contrasting algorithms for several pairs of
logics that differ only in the kind of transitivity supported.

Secondly, we present an algorithm that decides satisfiability of \alc extended
with transitive and inverse roles, role hierarchies, and functional
restrictions.  This algorithm can also be used for checking satisfiability and
subsumption with respect to general concept inclusion axioms (and thus cyclic
terminologies) because these axioms can be ``internalised''.  The fact that
our algorithm needs to deal only with transitive roles, instead of transitive
closure, leads to a 
lower degree of non-determinism, and experiments
indicate that the algorithm is well-suited for implementation.

Thirdly, we show that \alc extended with both transitive \emph{and} inverse
roles is still in \Pspace. The algorithm used to prove this result introduces
an enhanced blocking technique that should also provide useful efficiency
gains in implementations of more expressive DLs.

Fourthly, we investigate the limits of decidability for this family of DLs,
showing that relaxing the constraints we will impose on the kind of roles
allowed in number restrictions leads to the undecidability of all inference
problems.

Finally, we describe a range of optimisation techniques that can be
used to produce implementations of our algorithms that exhibit good
typical case performance.

\section{Preliminaries}

In this section, we present the syntax and semantics of the various DLs that are
investigated in subsequent sections. This includes the definition of inference
problems (concept subsumption and satisfiability, and both of these problems
with respect to terminologies) and how they are interrelated.

The logics we will discuss are all based on an extension of the well known DL
\alc~\cite{SSSm91} to include transitively closed primitive
roles~\cite{Sat95ashort}; we will call this logic \s due to its relationship
with the propositional (multi) modal logic
\Sfourm~\cite{Schi91}.\footnote{This logic has previously been called
  \alcRtrans, but this becomes too cumbersome when adding letters to represent
  additional features.} This basic DL is then extended in a variety of
ways---see Figure~\ref{fig:Sfamily} for an overview.

\begin{Definition}\label{syntax+semantics}
  Let $N_C$ be a set of \emph{concept names} and \Roles a set of \emph{role
  names} with transitive role names $\Rplus\subseteq \Roles$.
  The set of \si -{\em roles\/} is $\Roles \cup \{R^-\mid R\in\Roles \}$.  To
  avoid considering roles such as $R^{--}$, we define a function $\Inv$ on roles
  such that $\Inv(R) = R^-$ if $R$ is a role name, and $\Inv(R) = S$ if $R=S^-$.
  In the following, when speaking of roles, we refer to \si-roles, as our
  approach is capable of dealing uniformly with both role names and inverse
  roles.

  Obviously, a role $R$ is transitive iff $\Inv(R)$ is transitive.  We therefore
  define $\Tr$ to return $\mathrm{true}$ iff $R$ is a transitive role. More
  precisely, $\Tr(R) = \mathrm{true}$ (and we say that $R$ is transitive) iff
  $R\in \Rplus$ or $\Inv(R)\in\Rplus$.

  The set of \si-\emph{concepts} is the smallest set such that
\begin{enumerate}
\item every concept name is a concept, and,
\item if $C$ and $D$ are concepts and $R$ is an \si -role, then $(C\sqcap D)$,
  $(C\sqcup D)$, $(\neg C)$, $(\forall R.C)$, and $(\exists R.C)$ are also
  concepts.
\end{enumerate}
A \emph{role inclusion axiom} is of the form $R\sqsubseteq S$, where $R$ and $S$
are two roles, each of which can be inverse. A \emph{role hierarchy} is a finite
set of role inclusion axioms, and \shi is obtained from \si by allowing,
additionally, for a role hierarchy \R. The \emph{sub-role relation} $\sss$ is
the transitive-reflexive closure of $\sqsubseteq$ over $\R\cup\{
\Inv(R)\sqsubseteq \Inv(S)\mid R\sqsubseteq S \in \R\}$.

  \shiq is obtained from \shi by allowing, additionally, for qualified number
  restrictions~\cite{HollunderBaader-KR-91}, i.e., for concepts of the form
  $\atmostq{n}{R}{C}$ and $\atleastq{n}{R}{C}$, where $R$ is a \emph{simple}
  role, $C$ is a concept, and $n \in \N$.  A role is called \emph{simple} iff it
  is neither transitive nor has transitive sub-roles. \shin is the restriction
  of \shiq allowing only unqualified number restrictions (i.e., concepts of the
  form $\atmost{n}{R}$ and $\atleast{n}{R}$), while \shif represents a further
  restriction where, instead of arbitrary number restrictions, only
  \emph{functional restrictions} of the form $\atmost 1 R$ and their negation
  $\atleast 2 R$ may occur.
  
  An {\em interpretation\/} $\I = (\Delta^\I,\cdot^\I)$ consists of a set
  $\Delta^\I$, called the {\em domain\/} of $\I$, and a function $\cdot^\I$
  which maps every concept to a subset of $\Delta^\I$ and every role to a
  subset of $\Delta^\I\times\Delta^\I$ such that, for all concepts $C$, $D$,
  roles $R$, $S$, and non-negative integers $n$, the properties in
  Figure~\ref{fig:Sfamily} are satisfied, where $\sharp M$ denotes the
  cardinality of a set $M$. An interpretation satisfies a role hierarchy
  $\R$ iff $R^\I \subseteq S^\I$ for each $R \sqsubseteq S \in \R$; we denote
  this fact by $\I \models \R$ and say that $\I$ is a model of $\R$.

  A concept $C$ is called {\em satisfiable\/} with respect to a role hierarchy
  $\R$ iff there is some interpretation $\I$ such that $\I \models \R$ and
  $C^\I \neq \emptyset$.  Such an interpretation is called a {\em model of\/}
  $C$ w.r.t.\ $\R$. A concept $D$ {\em subsumes\/} a concept $C$ w.r.t.\ $\R$
  (written $C \sqsubseteq_{\R} D$) iff $C^\I \subseteq D^\I$ holds for each
  model $\I$ of $\R$.  For an interpretation $\I$, an individual $x \in
  \Delta^\I$ is called an {\em instance} of a concept $C$ iff $x\in C^\I$.

\end{Definition}
All DLs considered here are closed under negation, hence subsumption and
(un)satisfi\-ability w.r.t.\ role hierarchies can be reduced to each other: $C
\sqsubseteq_{\R} D$ iff $C\sqcap \neg D$ is unsatisfiable w.r.t.\ $\R$, and
$C$ is unsatisfiable w.r.t.\ $\R$ iff $C\sqsubseteq_{\R} A\sqcap \neg A$ for
some concept name $A$.

\begin{figure}[t!]
\begin{center}
{\setlength{\arraycolsep}{0em}
\begin{tabular}{|@{$\,$}l@{$\,$}|@{$\,$}c@{$\,$}|@{$\,$}c@{$\,$}|@{$\,$}c@{$\,$}|}
\hline
Construct Name & Syntax & Semantics & \\
\hline
atomic concept
 & $A$ & $A\ifunc \subseteq \deltai$ & \\
 atomic role
 & $R$ & $R\ifunc \subseteq \deltai \times \deltai$ & \\
 transitive role
 & $R \in \Rplus$ & $R\ifunc = (R\ifunc)^+$ & \\
 conjunction
 & $C \sqcap D$ & $C\ifunc \cap D\ifunc$ & \\
 disjunction
 & $C \sqcup D$ & $C\ifunc \cup D\ifunc$ & $\mathcal{S}$ \\
 negation
 & $\neg C$ & $\deltai\setminus C\ifunc$ & \\
 exists restriction & \some{R}{C} & $\{x \mid \exists y.\tuple{x}{y} \in R\ifunc
 \mbox{ and }y \in C\ifunc\}$ & \\
 value restriction &\all{R}{C}
 & $\{x \mid \forall y.\tuple{x}{y} \in R\ifunc \mbox{ implies }y \in C\ifunc\}$ & \\
\hline
role hierarchy
 & $R \sqsubseteq S$ & $R\ifunc \subseteq S\ifunc$ & $\mathcal{H}$ \\
\hline
inverse role
 & $R^-$ & $\{\tuple{x}{y} \mid \tuple{y}{x} \in R\ifunc\} $ & $\mathcal{I}$ \\
\hline
$\begin{array}{@{}l@{}}
 \mbox{number}\\
 \mbox{restrictions}
 \end{array}$
 & $\begin{array}{c}
    \atleast{n}{R} \\
    \atmost{n}{R}
    \end{array}$
 & $\begin{array}{l}
   \{x \mid \sharp \{y.\tuple{x}{y} \in R\ifunc\} \geqslant n\} \\
   \{x \mid \sharp \{y.\tuple{x}{y} \in R\ifunc\} \leqslant n\}
    \end{array}$
 & $\mathcal{N}$ \\
\hline
$\begin{array}{@{}l@{}}
 \mbox{qualifying number}\\
 \mbox{restrictions}
 \end{array}$
 & $\begin{array}{c}
    \atleastq{n}{R}{C} \\
    \atmostq{n}{R}{C}
    \end{array}$
 & $\begin{array}{l}
    \{x \mid \sharp \{y.\tuple{x}{y} \in R\ifunc \mbox{ and } y \in C\ifunc\} \geqslant n\} \\
    \{x \mid \sharp \{y.\tuple{x}{y} \in R\ifunc \mbox{ and }  y \in C\ifunc\} \leqslant n\}
    \end{array}$
 & $\mathcal{Q}$ \\
\hline
\end{tabular}}
\end{center}
\caption{Syntax and semantics of the \si family of DLs}\label{fig:Sfamily}
\end{figure}

\medskip
In \cite{KoTi90,Baad90b,Schi91,BBNNS93}, the \emph{internalisation} of
terminological axioms is introduced, a technique that reduces reasoning with
respect to a (possibly cyclic) \emph{terminology} to satisfiability of
concepts.  In \cite{Horrocks98c}, we saw how role hierarchies can be used for
this reduction. In the presence of inverse roles, this reduction must be
slightly modified.

\begin{Definition}
  
  A \emph{terminology} \Term is a finite set of general concept inclusion
  axioms, $\Term=\{C_1\sqsubseteq D_1,\ldots, C_n\sqsubseteq D_n\}$, where
  $C_i,D_i$ are arbitrary \shif-concepts. An interpretation $\I$ is said to be a
  \emph{model} of \Term iff $C_i\ifunc \subseteq D_i\ifunc$ holds for all $C_i
  \sqsubseteq D_i\in \Term$.  A concept $C$ is \emph{satisfiable} with respect
  to \Term iff there is a model $\I$ of \Term with $C\ifunc \not = \emptyset$.
  Finally, $D$ \emph{subsumes} $C$ with respect to \Term iff, for each model
  $\I$ of \Term, we have $C\ifunc \subseteq D\ifunc$.
\end{Definition}

The following lemma shows how general concept inclusion axioms can be
\emph{internalised} using a ``universal'' role $U$, a
transitive super-role of all roles occurring in \Term and their respective inverses.

\begin{Lemma}\label{lemma:terminologies}
  Let \Term be a terminology, $\R$ a role hierarchy, 
  and $C,D$ \shif-concepts, and let
  $$C_\Term:= \bigsqcap_{C_i\sqsubseteq D_i \in\Term}\neg C_i\sqcup D_i.$$
  \noindent  Let $U$ be a transitive role that does not occur in $\Term,C,D$, or $\R$. We
  set 
  $$\R_U := \R \cup \{ R\sqsubseteq U, \Inv(R)\sqsubseteq U \mid \text{$R$
  occurs in $\Term,C,D$, or $\R$} \}.$$
  Then $C$ is satisfiable w.r.t.\ \Term and $\R$ iff
  $C\sqcap C_\Term \sqcap \all{U}{C_\Term}$
  is satisfiable w.r.t.\ $\R_U$. Moreover, $D$ subsumes $C$ w.r.t.\ 
  \Term and $\R$ iff
  $C\sqcap\neg D\sqcap C_\Term \sqcap \all{U}{C_\Term}$
  is unsatisfiable w.r.t.\ $\R_U$.
\end{Lemma}

The proof of Lemma~\ref{lemma:terminologies} is similar to the ones that can be
found in \cite{Schi91,Baad90b}. Most importantly, it must be shown that, (a) if
a \shif-concept $C$ is satisfiable with respect to a terminology \Term and a
role hierarchy $\R$, then $C$, $\Term$, and $\R$ have a \emph{connected} model,
and (b) if $y$ is reachable from $x$ via a role path (possibly involving inverse
roles) in a model of $\Term$ and $\R_U$, then $\tuple{x}{y} \in U\ifunc$. These
are easy consequences of the semantics and the definition of $U$.

\begin{Theorem}\label{theorem:internal}
  Satisfiability and subsumption of \shif-concepts (resp.\ \shi-concepts)
  w.r.t.\ terminologies and role hierarchies are polynomially reducible to
  (un)satisfiability of \shif-concepts (resp.\ \shi-concepts) w.r.t.\ role
  hierarchies.
\end{Theorem}

\newcommand{\alcreg}{\ensuremath{\mathcal{ALC}_{\text{reg}}}\xspace}
\newcommand{\alcplus}{\ensuremath{\mathcal{ALC}_+}\xspace}
\newcommand{\alciplus}{\ensuremath{\mathcal{ALCI}_+}\xspace}
\newcommand{\alcireg}{\ensuremath{\mathcal{ALCI}_{\text{reg}}}\xspace}
\newcommand{\exptime}{\textsc{Exptime}\xspace}
\section{Blocking}
\label{sec:blocking}

\everypsbox{\scriptstyle}

The algorithms we are going to present for deciding satisfiability of \si- and
\shif-concepts use the tableaux method~\cite{HoNS90}, in which the
satisfiability of a concept $D$ is tested by trying to construct a model of $D$.
The model is represented by a tree in which nodes correspond to individuals and
edges correspond to roles. Each node $x$ is labelled with a set of concepts
$\Lab(x)$ that the individual $x$ must satisfy, and edges are labelled with 
(sets of) role names.

An algorithm starts with a single node labelled $\{D\}$, and proceeds by
repeatedly applying a set of \emph{expansion rules} that recursively decompose
the concepts in node labels, new edges and nodes being added as required in order
to satisfy $\some{R}{C}$ or $\Negfeat{F}$ concepts. The construction terminates either when none
of the rules can be applied in a way that extends the tree, or when the
discovery of obvious contradictions demonstrates that $D$ has no model.

In order to prove that such an algorithm is a sound and complete decision
procedure for concept satisfiability in a given logic, it is necessary to
demonstrate that the models it constructs are correct with respect to the
semantics, that it will always find a model if one exists, and that it always
terminates. The first two points can usually be dealt with by proving that the
expansion rules preserve satisfiability, and that in the case of
non-deterministic expansion (e.g., of disjunctions) all possibilities are
exhaustively searched. For logics such as \alc, termination is mainly due to the
fact that the expansion rules can only add new concepts that are strictly
smaller than the decomposed concept, so the model must stabilise when all
concepts have been fully decomposed. As we will see, this is no longer true in 
the presence of transitive roles.

\subsection{Transitive Roles vs. Transitive Closure}

We have argued that reasoning for logics with transitive roles is empirically
more tractable than for logics that allow for transitive closure of roles
\cite{Sat95ashort,Horrocks98c}. In this section we will give some justification
for that claim. The starting point for our investigations are the logics \sh
\cite{Horrocks98c} and \alcplus~\cite{Baad90b}, which extend \alc by transitive
roles and role hierarchies or transitive closure of roles respectively.
Syntactically, \alcplus is similar to \s, where, in addition to transitive and
non-transitive roles, the transitive closure $R^+$ of a role $R$ may appear in
existential and universal restrictions. Formally, $R^+$ is interpreted by
\[
(R^+)\ifunc = \bigcup_{i \in \N} (R\ifunc)^i, \quad \text{where } (R\ifunc)^i = 
\begin{cases}
  R\ifunc, & \text{if } i=1\\
  R\ifunc \circ (R\ifunc)^{i-1}, & \text{otherwise}
\end{cases}
\]

For both \sh and \alcplus, concept satisfiability is an \exptime-complete
problem. This result is easily derived from the \exptime-hardness proof for
\textsf{PDL} in \cite{FiLa79} and from the proof that \textsf{PDL} is in
\exptime in \cite{Prat79}.
Nevertheless, implementations of algorithms for \sh exhibit good performance in
realistic applications \cite{HoPa99} whereas, at the moment, this seems to be
more problematical for \alcplus. We believe that the main reason for this
discrepancy, at least in the case of tableau algorithm implementations, lies in
the different complexity of the blocking conditions that are needed to guarantee
the termination of the respective algorithms. In the following we are going to
survey the blocking techniques needed to deal with \sh and its subsequent
extensions to \shi and \shif. To underpin our claim that reasoning with
transitive roles empirically leads to more efficient implementations than for
transitive closure, we will also present the blocking techniques used to deal
with transitive closure. These are more complicated and introduce a larger
degree in non-determinism into the tableaux algorithms, leading to inferior
performance of implementations.

\subsection{Blocking for \s and \sh}

Termination of the expansion process of a tableaux algorithm is not
guaranteed for logics that include transitive roles, as the expansion rules
can introduce new concepts that are the same size as the decomposed concept.
In particular, $\all{R}{C}$ concepts, where $R$ is a transitive role, are
dealt with by propagating the whole concept across $R$-labelled
edges~\cite{Sat95ashort}. For example, given a node $x$ labelled $\{C,
\some{R}{C}, \all{R}{(\some{R}{C})}\}$, where $R$ is a transitive role, the
combination of the $\some{R}{C}$ and $\all{R}{(\some{R}{C})}$ concepts would
cause a new node $y$ to be added to the tree with a label identical to that
of $x$.
The expansion process could then be repeated indefinitely.

This problem can be dealt with by \emph{blocking}: halting the expansion process
when a \emph{cycle} is detected~\cite{Baad90b,Buchheit93}. For logics without inverse
roles, the general procedure is to check the label of each new node $y$, and if
it is a \emph{subset}~\cite{Baader96a} of the label of an ancestor node $x$,
then no further expansion of $y$ is performed: $x$ is said to block $y$. The
resulting tree corresponds to a cyclical model in which $y$ is identified with
$x$.

To deal with the transitive closure of roles, tableaux algorithms proceed by
non-deterministically expanding a concept $\some {R^+} C$ to either $\some R C$
or $\some R {\some {R^+} C}$. Again, since the size of concepts along a path in
the tree may not decrease, blocking techniques are necessary to guarantee
termination. An adequate blocking condition for \alcplus is identical as for
\sh, but one has to distinguish between \emph{good} and \emph{bad} cycles.
Consider the following concept:
\[
D = \some {R^+} A \sqcap \all {R^+} {\neg A} \sqcap \neg A
\]
While $D$ is obviously not satisfiable, a run of a tableaux algorithm might
generate the following tableau in which node $y$ is blocked by node $x$
without generating any obvious contradictions.
\[
\psset{nodesep=3pt,linewidth=.5pt}
\begin{array}{l}
  \rnode{a}{\bullet}_x \; \some {R^+} A, \; \all {R^+} {\neg A}, \; \some R
  {\some {R^+} A}, \; \neg A\\[6ex] 
  \rnode{b}{\bullet}_y \; \some {R^+} A, \; \all {R^+} {\neg A}, \; \some R {\some
    {R^+} A}, \; \neg A
\end{array}
\ncline{->}{a}{b}\Aput{R}
\]
The problem is that $\some {R^+} A$ has always been expanded to $\some R
{\some {R^+} A}$, postponing the satisfaction of $A$ a further step.
To obtain a correct tableaux algorithm for \alcplus, the blocking
condition must include a check to ensure that each concept $\some {R^+} C$ appearing in such
a cycle is expanded to $\some R C$ somewhere in the cycle. Such cycles
are called \emph{good} cycles, whereas cycles in which ${\some {R^+} C}$ has
always been expanded to $\some R {\some {R^+} C}$ are called \emph{bad}
cycles. A valid model may only contain good cycles.

Summing up, using transitive closure instead of transitive roles has a twofold
impact on the empirical tractability: (a) in blocking situations, good cycles
have to be distinguished from bad ones, and (b) the non-deterministic
expansion of concepts of the form $\some {R^+} C$ increases the size
of the search space.

\subsection{Adding Inverse Roles}
\label{sec:dynblk}

Blocking is more problematical when inverse roles are added to the logic, and
a key feature of the algorithms presented in~\cite{HoSat98c} was the
introduction of a \emph{dynamic blocking} strategy. Besides using label
equality instead of subset, this strategy allowed blocks to be established,
broken, and re-established. With inverse roles the blocking condition has to
be considered more carefully because roles are now bi-directional, and
additional concepts in $x$'s label could invalidate the model with respect to
$y$'s predecessor. This problem can be overcome by allowing a node $x$ to be
blocked by one of its ancestors $y$ if and only if they were labelled with the
same sets of concepts.

Dealing with inverse roles is even more complicated in the presence of
transitive closure. As an example consider the following concept:
\begin{align*}
  D & = \neg A \sqcap \some R {\some {R^+} C}\\
  C & = \all {R^-} {( \all {R^-} A)}
\end{align*}

Fig.~\ref{fig:trans-inv} shows two possible tableau expansions of the concept
$D$. Continuing the expansion of the left hand tree will necessarily lead to a
clash when concept $C \in \Lab(z)$ is expanded as this will lead to
both $A$ and $\neg A$ appearing in $\Lab(x)$. The right hand tree is
also invalid as it contains a bad cycle:
$\Lab(y) = \Lab(z)$ but $\some {R^+} D$ has always been expanded to
$\some R \some {R^+} D$. Nevertheless, $C$ is satisfiable, as
would be shown by continuing the expansion of the right hand path for
one more step.

\begin{figure}
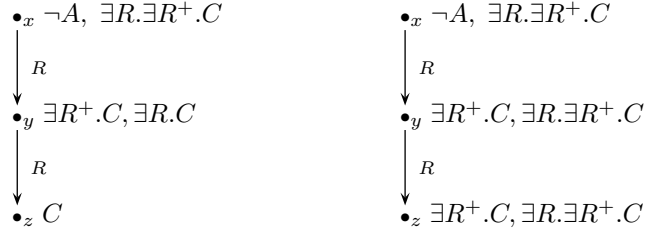

  \psset{nodesep=3pt,linewidth=.5pt}
  \begin{center}
      $\begin{array}{c@{\hspace{2cm}}c}
        \begin{array}{l}
          \rnode{d}{\bullet}_x \; \neg A, \; \some R {\some {R^+} C}\\[6ex]
          \rnode{e}{\bullet}_y \; \some {R^+} C, \some R C\\[6ex]
          \rnode{f}{\bullet}_z \; C
        \end{array}
        &
        \begin{array}{l}
          \rnode{a}{\bullet}_x \; \neg A, \; \some R {\some {R^+} C}\\[6ex]
          \rnode{b}{\bullet}_y \; \some {R^+} C, \some R {\some {R^+} C}\\[6ex]
          \rnode{c}{\bullet}_z \; \some {R^+} C, \some R {\some {R^+} C}
        \end{array}
        \ncline{->}{a}{b}\Aput{R}  
        \ncline{->}{b}{c}\Aput{R}
        \ncline{->}{d}{e}\Aput{R}
        \ncline{->}{e}{f}\Aput{R}
      \end{array}
      $
  \end{center}
  \caption{Dynamic blocking fails in the presence of transitive closure.}
  \label{fig:trans-inv}
\end{figure}

In \cite{DeGiaMass98}, a solution to this problem for \textsf{CPDL}, a strict
superset of \alciplus (\alcplus plus inverse roles) is presented. The solution
consists of an additional expansion rule called the \emph{look behind analytical
  cut}. This rule employs exhaustive non-deterministic guessing to make the past
of each node in the tree explicit in the labelling of that node: if $y$ is an
$R$-successor of a node $x$, then $\some {R^-} C$ or $\all {R^-} \neg
C$ is added non-deterministically to the label of $y$ for each concept $C$
that may appear during the expansion process. Obviously, this leads to
a further large increase in the size of the search space, with a
correspondingly large adverse impact on empirical
tractability. Experience with this kind of exhaustive guessing leads us 
to believe that an implementation of such an algorithm would be
disastrously inefficient. The non-existence of implementations
for \alciplus or \textsf{CPDL} might be taken to support this view.

\subsection{Pair-wise Blocking}
\label{sec:pairblk}

Further extending the logic \shi to \shif by adding functional restrictions
(concepts of the form \feat{R}, meaning that an individual can be related to
at most one other individual by the role $R$) introduces new problems
associated with the fact that the logic no longer has the finite model
property. This means that there are concepts that are satisfiable but for
which there exists no finite model. An example of such a concept is
$$\neg C \sqcap \some{F^-}{(C \sqcap \feat{F})} \sqcap
\all{R^-}{(\some{F^-}{(C\sqcap \feat{F} )}) },$$
where $R$ is a transitive role and $F \sqsubseteq R$. Any model of
this concept must contain an infinite sequence of individuals, each
related to a single successors by an $F^-$ role, and each satisfying
$C \sqcap \some{F^-}{C}$, the $\some{F^-}{C}$ term being propagated
along the sequence by the transitive super-role $R$.  Attempting to
terminate the sequence in a cycle causes the whole sequence to collapse
into a single node due to the functional restrictions \feat{F}, and
this results in a contradiction as both $C$ and $\neg C$ will be in
that node's label.

In order to deal with infinite models---namely to have an algorithm that
terminates correctly even if the input concept has only infinite models---a more
sophisticated \emph{pair-wise} blocking strategy was introduced
in~\cite{HoSat98c}, and soundness was proved by demonstrating that a blocked
tree always has a corresponding infinite model.\footnote{This is not to say that
it may not also have a finite model.}

The only known algorithm that is able to deal with the combination of
transitive closure, inverse roles, and functional restrictions on roles relies
on an elaborate polynomial reduction to a \textsf{CPDL} terminology
\cite{DeLe94c}, and the capability of \textsf{CPDL} to internalise the
resulting general terminological axioms. The large number and the nature of
the axioms generated by this reduction make it very unlikely that an
implementation with tolerable runtime behaviour will ever emerge.

\section{Reasoning for \si Logics} \label{sec:alchfi}

In this section, we present two tableaux algorithms: the first decides
satisfiability of \shif-concepts, and can be used for all \shif reasoning
problems (see Theorem~\ref{theorem:internal}); the second decides satisfiability
(and hence subsumption) of \si-concepts in \Pspace. In this paper we only sketch
most of the proofs. For details on the \shif-algorithm, please refer to
\cite{HoSat98c}, for details on the \si- and \sin-algorithm, please refer to
\cite{Horrocks98k}.

The correctness of the algorithms can be proved by showing that they create a
\emph{tableau} for a concept iff it is satisfiable.

For ease of construction, we assume all concepts to be in \emph{negation
  normal form} (NNF), that is, negation occurs only in front of concept names.
Any \shif-concept can easily be transformed to an equivalent one in NNF by
pushing negations inwards~\cite{HoNS90}.

\begin{Definition} \label{def:alchr2}
  Let $D$ be a \shif-concept in NNF, $\R$ a role hierarchy, and $\Roles_D$
  the set of roles occurring in $D$ together with their inverses, and
  $\mathit{sub}(D)$ the subconcepts of $D$. Then $T =
  (\mathbf{S},\Lab,\Edges)$ is a \emph{tableau} for $D$ w.r.t.\ $\R$ iff
  $\mathbf{S}$ is a set of individuals, $\Lab:\mathbf{S} \rightarrow
  2^{\mathit{sub}(D)}$ maps each
  individual to a set of concepts, 
  $\Edges:\Roles_D \rightarrow 2^{\mathbf{S} \times \mathbf{S}}$ maps each role
  to a set of pairs of individuals, and there is some individual $s \in
  \mathbf{S}$ such that $D \in \Lab(s)$. Furthermore, for all $s,t \in
  \mathbf{S}$, $C,E\in \mathit{sub}(D)$, and $R,S \in \Roles_D$, it holds that:
  \begin{enumerate}\setlength{\itemsep}{0ex}
  \item if $C \in \Lab(s)$, then $\neg C \notin \Lab(s)$,
    
  \item if $C \sqcap E \in \Lab(s)$, then $C\in\Lab(s)$ and ${E\in\Lab(s)}$,
    
  \item if $C \sqcup E \in \Lab(s)$, then $C \in \Lab(s)$ or $E \in \Lab(s)$,
    
  \item if $\all{R}{C} \in \Lab(s)$ and $\tuple{s}{t} \in \Edges(R)$, then $C \in
    \Lab(t)$,
  
  \item if $\some{R}{C} \in \Lab(s)$, then there is some $t \in \mathbf{S}$ such
    that $\tuple{s}{t} \in \Edges(R)$ and $C \in \Lab(t)$,
  \item if $\all{S}{C} \in \Lab(s)$ and $\tuple{s}{t} \in \Edges(R)$ for
    some $R\sss S$ with $\Tr(R)$, then $\all{R}{C} \in \Lab(t)$,
    
  \item $\tuple{s}{t}\in \Edges(R)$ iff $\tuple{t}{s}\in \Edges(\Inv(R))$.

  \item if $\tuple{x}{y}\in \Edges(R)$ and $R\sss S$, then $\tuple{x}{y}\in
    \Edges(S)$,
  
  \item if $\atmost{1}{R} \in \Lab(s)$, then $\sharp\{t\mid\tuple{s}{t} \in
    \Edges(R) \} \leq 1$, and
    
  \item if $\atleast{2}{R} \in \Lab(s)$, then $\sharp\{t\mid\tuple{s}{t} \in
    \Edges(R) \} \geq 2$.
  \end{enumerate}
  
  Tableaux for \si-concepts are defined analogously and must satisfy Properties
  1-7, where, due to the absence of a role hierarchy, $\sss$ is the identity.
\end{Definition}

Due to the close relationship between models and tableaux, the following lemma
can be easily proved by induction on the structure of concepts.  As a
consequence, an algorithm that constructs (if possible) a tableau for an input
concept is a decision procedure for satisfiability of concepts.

\begin{Lemma}\label{lemma:shin-tabl}
  A \shif-concept (resp.\ \si-concept) $D$ is satisfiable w.r.t.\ a role hierarchy
  $\R$ iff $D$ has a tableau w.r.t.\ $\R$. 
\end{Lemma}

\subsection{Reasoning in \shif}\label{sec:shin-algo}

In the following, we give an algorithm that, given a \shif-concept $D$,
decides the existence of a tableaux for $D$. We implicitly assume an arbitrary
but fixed role hierarchy $\R$.

\begin{Definition}\label{def:shin-algo}
  A \emph{completion tree} for a \shif-concept $D$ is a tree where each node $x$
  of the tree is labelled with a set $\Lab(x) \subseteq \textit{sub}(D)$ and
  each edge $\tuple{x}{y}$ is labelled with a set $\Lab(\tuple{x}{y})$ of (possibly inverse)
  roles occurring in $\textit{sub}(D)$.
  
  Given a completion tree, a node $y$ is called an $R$-\emph{successor} of a
  node $x$ iff $y$ is a successor of $x$ and $S\in \Lab(\tuple{x}{y})$ for some
  $S$ with $S\sss R$.  A node $y$ is called an $R$\emph{-neighbour} of $x$ iff
  $y$ is an $R$-successor of $x$, or if $x$ is an $\Inv(R)$-successor of $y$.
  Predecessors and ancestors are defined as usual.
  
  A node is \emph{blocked} iff it is directly or indirectly blocked.  A node $x$
  is \emph{directly blocked} iff none of its ancestors are blocked, and it has
  ancestors $x'$, $y$ and $y'$ such that
  \begin{enumerate}\setlength{\itemsep}{0ex}
    \item $x$ is a successor of $x'$ and $y$ is a successor of $y'$ \emph{and}
    \item $\Lab(x) = \Lab(y)$ and $\Lab(x') = \Lab(y')$ \emph{and}
    \item $\Lab(\tuple{x'}{x}) = \Lab(\tuple{y'}{y})$.
  \end{enumerate}
  In this case we will say that $y$ blocks $x$.
  
  A node $y$ is \emph{indirectly blocked} iff one of its ancestors is blocked,
  or---in order to avoid wasted expansion after an application of the
  $\leqslant$-rule---
  it is a successor of a node $x$ and $\Lab(\tuple{x}{y}) = \emptyset$.
  
  For a node $x$, $\Lab(x)$ is said to contain a {\em clash} iff $\{A,\neg A\}
  \subseteq \Lab(x)$ or $\{\atleast{2}{R}, \atmost{1}{S}\} \subseteq \Lab(x)$
  for roles $R \sss S$. A completion tree is called {\em
    clash-free} iff none of its nodes contains a clash; it is called
  {\em complete} iff none of the expansion rules in
  Figure~\ref{table:alchfi} is applicable.
  
  For a \shif-concept $D$ in NNF, the algorithm starts with a completion tree
  consisting of a single node $x$ with $\Lab (x)=\{D\}$. It applies the
  expansion rules, stopping when a clash occurs, and answers ``$D$ is
  satisfiable'' iff the completion rules can be applied in such a way that they
  yield a complete and clash-free completion tree.
\end{Definition}

\begin{figure}[t!]
  \begin{center}
    \begin{tabular}{@{ }l@{ }l@{ }l@{}}
      $\sqcap$-rule: & if \hfill 1. & $C_1 \sqcap C_2 \in \Lab(x)$, $x$ is not
      indirectly blocked, and \\
      & \hfill 2. & $\{C_1,C_2\} \not\subseteq \Lab(x)$
                                \\ & then &
      $\Lab(x) \longrightarrow \Lab(x) \cup \{C_1,C_2\}$ 
      \\[0.5ex]
      
      $\sqcup$-rule: & if \hfill 1. & $C_1 \sqcup C_2 \in \Lab(x)$, $x$ is not
      indirectly blocked, and \\
      & \hfill 2. &$\{C_1,C_2\} \cap \Lab(x) = \emptyset$ 
                                \\ & then, & 
        for some $C\in \{C_1,C_2 \}$,   $\Lab(x)
      \longrightarrow \Lab(x) \cup \{C\}$  \\[0.5ex]
      
      $\exists$-rule: & if \hfill 1. & $\some{S}{C} \in \Lab(x)$, $x$ is not
      blocked, and \\ 
      & \hfill 2. & $x$ has no $S$-neighbour $y$ with $C\in \Lab(y)$\\
      & then & 
       create a new node  $y$  with \\
       & &$\Lab(\tuple{x}{y})=\{S\}$ and $\Lab(y)=\{C\}$ \\[0.5ex]
      
      $\forall$-rule: & if \hfill 1. & $\all{S}{C} \in \Lab(x)$, $x$ is not
      indirectly blocked, and \\
      & \hfill 2. &there is an $S$-neighbour $y$ of $x$ with $C \notin \Lab(y)$\\
      & then &  
      $\Lab(y) \longrightarrow \Lab(y) \cup \{C\}$ \\[0.5ex]

      $\forall_+'$-rule: & if \hfill 1. & $\all{S}{C} \in
      \Lab(x)$,  $x$ is not indirectly blocked,\\ 
      & \hfill 2. &there is some $R$ with $\Tr(R)$ and $R\sss S$,  and\\
      &\hfill 3. & $x$ has an $R$-neighbour $y$ with $\all{R}{C} \notin
      \Lab(y)$ \\
      & then &  
       $\Lab(y) \longrightarrow \Lab(y) \cup
      \{\all{R}{C}\}$ \\[0.5ex] 
      
      $\geqslant$-rule: & if \hfill 1. & $\Negfeat{R} \in \Lab(x)$, $x$ is not
      blocked,  and \\
      & \hfill 2. &there is no $R$-neighbour $y$ of $x$ with $A \in \Lab(y)$\\
      & then & create two new nodes $y_1$, $y_2$ with\\
      & & $\Lab(\tuple{x}{y_1})=\Lab(\tuple{x}{y_2})=\{R\}$,\\
      & & $\Lab(y_1)= \{A\}$ and $\Lab(y_2)=
      \{\neg A\}$ \\[0.5ex] 
      $\leqslant$-rule: & if \hfill 1. & $\feat{R} \in \Lab(x)$,  $x$ is not
      indirectly blocked,\\  
      & \hfill 2. &$x$ has two $R$-neighbours $y$ and $z$ s.t.\ $y$ is not an ancestor of $z$,  \\
      &then & 1.\ $\Lab(z) \longrightarrow \Lab(z) \cup \Lab(y)$ and\\
      & & 2.\ 
      \begin{tabular}[t]{@{}l@{}}
        if $z$ is an ancestor of $y$ \\
        then 
        $\Lab (\tuple{z}{x}) \longrightarrow  \Lab (\tuple{z}{x}) \cup 
        \Inv(\Lab (\tuple{x}{y}))$\\
        else 
        $\Lab (\tuple{x}{z}) \longrightarrow  \Lab (\tuple{x}{z}) \cup 
        \Lab (\tuple{x}{y})$
      \end{tabular}\\
      & & 3.\ $\Lab(\tuple{x}{y}) \longrightarrow \emptyset$
  \end{tabular}

\caption{The tableaux expansion rules for
\shif}\label{table:alchfi} \vspace{-0.6cm}
\end{center}
\end{figure}

The soundness and completeness of the tableaux
algorithm is an immediate consequence of Lemmas~\ref{lemma:shin-tabl} and
\ref{lemma:shin-algo-correct}.

\begin{Lemma} \label{lemma:shin-algo-correct}
  Let $D$ be an \shif-concept.
  \begin{enumerate}
  \item The tableaux algorithm terminates when started with $D$.
  \item If the expansion rules can be applied to $D$ such
    that they yield a complete and clash-free completion tree, then $D$ has a
    tableau.
  \item If $D$ has a tableau, then the expansion rules can be applied to
    $D$ such that they yield a complete and clash-free completion
    tree.
  \end{enumerate}
\end{Lemma}

Before we sketch the ideas of the proof, we will discuss the different expansion
rules and their correspondence to the language constructors. 

The $\sqcap$-, $\sqcup$-, $\exists$- and $\forall$-rules are the standard
\alc tableaux rules \cite{SSSm91}. The $\forall_+$-rule is used to
handle transitive roles, where the $\sss$-clause deals with the role
hierarchy.
See \cite{HoSat98c} for details.

The functional restriction rules merit closer consideration.  In order to
guarantee the satisfaction of a $\atleast 2 R$-constraint, the $\geqslant$-rule
creates two successors and uses a fresh atomic concept $A$ to prohibit
identification of these successors by the $\leqslant$-rule.
If a node $x$ has two or more $R$-neighbours and contains a functional
restriction $\atmost{1}{R}$, then the $\leqslant$-rule merges two of the
neighbours \emph{and} also merges the edges connecting them with $x$. Labelling
edges with sets of roles allows a single node to be both an $R$ and
$S$-successor of $x$ even if $R$ and $S$ are not comparable by $\sss$.
Finally, contradicting functional restrictions are taken care of by the
definition of a clash.

We now sketch the main ideas behind the proof of
Lemma~\ref{lemma:shin-algo-correct}:

\textbf{1. Termination:} Let $m = \Card{\textit{sub}(D)}$ and $n =
\Card{\Roles_D}$.  Termination is a consequence of the following properties of
the expansion rules:

(a) The expansion rules never remove nodes from the tree or concepts from node
labels. Edge labels can only be changed by the $\leqslant$-rule which either
expands them or sets them to $\emptyset$; in the latter case the node below the
$\emptyset$-labelled edge is blocked.
(b) Successors are only generated for concepts of the form \some{R}{C} and
\atleast{2}{R}. For a node $x$, each of these concepts triggers the generation
of at most two successors. If for one of these successors $y$ the
$\leqslant$-rule subsequently causes $\Lab(\tuple{x}{y})$ to be changed to
$\emptyset$, then $x$ will have some $R$-neighbour $z$ with $\Lab(z) \supseteq
\Lab(y)$. This, together with the definition of a clash, implies that the
concept that led to the generation of $y$ will not trigger another rule
application. Obviously, the out-degree of the
tree is bounded by $2m$. 
(c) Nodes are labelled with non-empty subsets of $\textit{sub}(D)$ and edges
with subsets of $\Roles_D$, so there are at most $2^{2mn}$ different possible
labellings for a pair of nodes and an edge.  Therefore, on a path of length at
least $2^{2mn}$ there must be 2 nodes $x,y$ such that $x$ is directly blocked by
$y$.  Since a path on which nodes are blocked cannot become longer, paths are of
length at most $2^{2mn}$.

\textbf{2. Soundness:} A complete and clash-free tree \Tree for $D$ induces the
existence of a tableaux $T = (\mathbf{S},\Lab,\Edges)$ for $D$ as follows.
Individuals in $\mathbf{S}$ correspond to {\em paths} in $\Tree$ from the root
node to some node that is not blocked.  Instead of going to a directly blocked
node, these paths jump back to the blocking node, which yields paths of
arbitrary length.  Thus, if blocking occurs, this construction yields an
infinite tableau. This rather complicated tableau construction is necessary due
to the presence of functional restrictions; its validity is ensured by the
blocking condition, which considers both the blocked node and its predecessor.

\textbf{3. Completeness:} A tableau $T= (\mathbf{S},\Lab,\Edges)$ for $D$ can be
used to ``steer'' the application of the non-deterministic $\sqcup$-
and $\leqslant$-rules in a way that yields a complete and clash-free tree.

The following theorem is an immediate consequence of
Lemma~\ref{lemma:shin-algo-correct}, Lemma~\ref{lemma:shin-tabl}, and
Lemma~\ref{lemma:terminologies}.

\begin{Theorem}\label{theorem:dec-shin}
  The tableaux algorithm is a decision procedure for the satisfiability and
  subsumption of \shif-concepts with respect to terminologies and role
  hierachies.
\end{Theorem}

\newcommand{\pspace}{\textsc{PSpace}\xspace}

\subsection{A \pspace-algorithm for \si}

To obtain a \pspace-algorithm for \si, the \shif algorithm is modified as
follows: (a) As \si does not allow for functional restrictions, the $\geqslant$-
and the $\leqslant$-rule can be omitted; blocking no longer involves two pairs
of nodes with identical labels but only two nodes with ``similar'' labels. (b)
Due to the absence of role hierarchies, edge labels can be restricted to roles
(instead of sets of roles).  (c) To obtain a \pspace algorithm, we employ a
refined blocking strategy which necessitates a second label $\BLab$ for each
node. This blocking technique, while discovered independently, is based on ideas
similar to those used in \cite{spaan_e:1993a} to show that satisfiability for
$\mathbf{K4}_{\mathbf{t}}$ can be decided in \pspace.%
\footnote{The modal logic $\mathbf{K4}_{\mathbf{t}}$ is a syntactic variant of
\si with only a single transitive role name.}
In the following, we will describe and motivate this blocking technique;
detailed proofs as well as a similar result for \sin can be found in
\cite{Horrocks98k}.

Please note that naively using a cut rule does not yield a PSpace algorithm: a
cut rule similar to the \emph{look behind analytical cut} presented
in~\cite{DeGiaMass98} (non-deterministically) guesses which constraints will be
propagated ``up'' the completion tree by universal restrictions on inverted
roles. For \si, this technique may lead to paths of exponential length due to
equality blocking. A way to avoid these long paths would be to stop the
investigation of a path at some polynomial bound. However, to prove the
correctness of this approach, it would be necessary to establish a
``short-path-model'' property similar to Lemma~\ref{lemma:poly_paths}.
Furthermore, we believe that our algorithm is better suited for an
implementation since it makes less use of ``don't-know'' non-determinism. This
also distinguishes our approach from the algorithm presented
\cite{spaan_e:1993a}, which is not intended to form the basis for an efficient
implementation.

\begin{Definition}\label{def:si-algo}
  A \emph{completion tree} for a \si concept $D$ is a tree where each node $x$
  of the tree is labelled with two sets $\BLab(x) \subseteq \Lab(x) \subseteq
  \textit{sub}(D)$ and each edge $\tuple{x}{y}$ is labelled with a (possibly
  inverse) role $\Lab(\tuple{x}{y})$ occurring in $\textit{sub}(D)$.
  
  $R$-neighbours, -successors, and -predecessors are defined as in
  Definition~\ref{def:shin-algo}. Due to the absence of role
  hierarchies, $\sss$ is  the identity on $\Roles_D$.
  
  A node $x$ is \emph{blocked} iff, for an ancestor $y$, $y$ is blocked or
  $$\BLab(x) \subseteq \Lab(y)\quad  \text{ and }\quad \Lab(x)/\Inv(S) =
  \Lab(y)/\Inv(S),$$
  where $x'$ is the predecessor of $x$,
  $\Lab(\tuple{x'}{x})=S$, and 
  $\Lab(x)/\Inv(S) = \{ \all{\Inv(S)}{C} \in \Lab(x)\}$.

  For a node $x$, $\Lab(x)$ is said to contain a \emph{clash} iff $\{ A, \neg A
  \} \subseteq \Lab(x)$. 
  A completion tree to which none of the expansion
  rules given in Figure~\ref{table:alci} is applicable is called \emph{complete}.
  
  For an \si-concept $D$, the algorithm starts with a
  completion tree consisting of a single node $x$ with
  $\BLab(x) = \Lab(x) =
  \{D\}$. It applies the expansion rules in Figure~\ref{table:alci}, stopping when
  a clash occurs, and answers ``$D$ is satisfiable'' iff the completion rules
  can be applied in such a way that they yield a complete and clash-free
  completion tree.
\end{Definition}

\begin{figure}
  \begin{center}
    \begin{tabular}{l@{$\;$}l@{$\;$}l}
      $\sqcap$-rule: & if \hfill 1. & $C_1 \sqcap C_2 \in \Lab(x)$ and \\ &
      \hfill 2. & $\{C_1,C_2\} \not\subseteq \Lab(x)$ \\ & then & $\Lab(x)
      \longrightarrow \Lab(x) \cup \{C_1,C_2\}$ \\

      $\sqcup$-rule: & if \hfill 1. & $C_1 \sqcup C_2 \in \Lab(x)$ and \\ &
      \hfill 2. & $\{C_1,C_2\} \cap \Lab(x) = \emptyset$ \\ & then & $\Lab(x)
      \longrightarrow \Lab(x) \cup \{C\}$ for some $C\in \{C_1,C_2 \}$\\
      $\forall$-rule: & if \hfill 1. & $\all{S}{C} \in \Lab(x)$ and \\ & \hfill
      2. &there is an $S$-successor $y$ of $x$ with $C \notin \BLab(y)$ \\ & then &
      $\Lab(y) \longrightarrow \Lab(y) \cup \{C\}$ and \\
      & & $\BLab(y) \longrightarrow
      \BLab(y) \cup \{C\}$ or\\ & \hfill 2'.&there is an $S$-predecessor $y$ of $x$
      with $C \notin \Lab(y)$ \\ & then & $\Lab(y) \longrightarrow \Lab(y) \cup
      \{C\}$. \\
      $\forall_+$-rule: & if \hfill 1. & $\all{S}{C} \in \Lab(x)$ and $\Tr(S)$
      and \\ & \hfill 2. &there is an $S$-successor\ $y$ of $x$ with $\all{S}{C}
      \notin \BLab(y)$ \\ & then & $\Lab(y) \longrightarrow \Lab(y) \cup
      \{\all{S}{C}\}$ and \\
      & & $\BLab(y) \longrightarrow \BLab(y) \cup
      \{\all{S}{C}\}$ or \\ & \hfill 2'. &there is an $S$-predecessor\ $y$ of $x$
      with $\all{S}{C} \notin \Lab(y)$ \\ & then & $\Lab(y) \longrightarrow
      \Lab(y) \cup \{\all{S}{C}\}$. \\
      $\exists$-rule: & if \hfill 1. & $\some{S}{C} \in \Lab(x)$, $x$ is not
      blocked and no other rule \\ && is applicable to any of its ancestors, and
      \\ & \hfill 2. & $x$ has no $S$-neighbour $y$ with $C\in \Lab(y)$ \\ &
      then & create a new node $y$ with \\
      & & $\Lab(\tuple{x}{y})=S$ and $\Lab(y)=
      \BLab(y) = \{C\}$ \\
    \end{tabular}
    \caption{Tableaux expansion rules for \si}
    \label{table:alci}\vspace{-0.5cm}
  \end{center}
\end{figure}

As for \shif, correctness of the algorithm is proved by first showing that a
\si-concept is satisfiable iff it has a tableau, and next proving the
\si-analogue of Lemma~\ref{lemma:shin-algo-correct}.
\begin{Theorem}\label{theo:si-algo}
  The tableaux algorithm is a decision procedure for satisfiability and
  subsumption of \si-concepts.
\end{Theorem}

The dynamic blocking technique for \si and \shi described in Section~\ref{sec:blocking}, which is
based on label equality, may lead to completion trees with exponentially long
paths because there are exponentially many possibilities to label sets on such
a path.  Due to the non-deterministic $\sqcup$-rule, these exponentially many
sets may actually occur.

This non-determinism is not problematical for $\mathcal{S}$ because disjunctions
need not be completely decomposed to yield a subset-blocking situation.  For an
optimal \si algorithm, the additional label $\BLab$ was introduced to enable a
sort of subset-blocking which is independent of the $\sqcup$-non-determinism.
Intuitively, $\BLab(x)$ is the restriction of $\Lab(x)$ to those non-decomposed
concepts that $x$ must satisfy, whereas $\Lab(x)$ contains boolean
decompositions of these concepts as well as those that are imposed by value
restrictions in descendants.  If $x$ is blocked by $y$, then all concepts in
$\BLab(x)$ are eventually decomposed in $\Lab(y)$ (if no clash occurs). However,
in order to substitute $x$ by $y$, $x$'s constraints on predecessors must be at
least as strong as $y$'s; this is taken care of by the second blocking
condition.

Let us consider a path $x_1,\dots,x_n$ where all edges are labelled $R$ with
$\Tr(R)$, the only kind of paths along which the length of the longest concept
in the labels might not decrease. If no rules can be applied, we have
$\Lab(x_{i+1})/\Inv(R) \subseteq \Lab(x_{i})/\Inv(R)$ and $\BLab(x_i)\subseteq
\BLab(x_{i+1}) \cup\{C_i\}$ (where $\some{R}{C_i}$ triggered the generation of
$x_{i+1}$). This limits the number of labels and guarantees blocking after
a polynomial number of steps.

\begin{Lemma}\label{lemma:poly_paths}
The paths of a completion tree for a concept $D$ have a length of at most $m^4$
where $m = |\textit{sub}(D)|$.
\end{Lemma}

Finally, a slight modification of the expansion rules given in
Figure~\ref{table:alci} yields a \pspace algorithm.
This modification is
necessary because the original algorithm must keep the whole completion tree in
its memory---which needs exponential space even though the length of its paths
is polynomially bounded. The original algorithm may not forget about branches
because restrictions which are pushed \emph{upwards} in the tree might make it
necessary to revisit paths which have been considered before. We solve
this problem as follows:

Whenever the $\forall$- or the $\forall_+$-rule is applied to a node $x$ and its
\emph{predecessor} $y$ (Case 2' of these rules), we delete all successors of $y$
from the completion tree. While this makes it necessary to restart the
generation of successors for $y$, it makes it possible to implement the
algorithm in a depth-first manner which facilitates the re-use of space.

This modification does not affect the proof of soundness and completeness for
the algorithm, but we have to re-prove termination \cite{Horrocks98k}
as it relied on the fact that we never removed any nodes from the
completion tree.  Summing up we get:
\begin{Theorem}
  The modified algorithm is a \pspace decision procedure for satisfiability
  and subsumption of \si-concepts.
\end{Theorem}

\newcommand{\shinplus}{\ensuremath{\mathcal{SHIN}^+}\xspace}
\newcommand{\shnplus}{\ensuremath{\mathcal{SHN}^+}\xspace}
\newcommand{\shn}{\ensuremath{\mathcal{SHN}}\xspace}

\section{The Undecidability of Unrestricted \shn}

In \cite{Horrocks99j} we describe an algorithm for \shiq based on the
\shif-algorithm already presented.  Like earlier DLs that combine a hierarchy
of (transitive and non-transitive) roles with some form of number
restrictions~\cite{HoSat98c,Horrocks98k} and \shif, the DL \shiq allows only
\emph{simple} roles in number restrictions. The justification for this
limitation has been partly on the grounds of a doubtful semantics (of
transitive functional roles) and partly to simplify decision procedures. In
this section we will show that, even for the simpler \shn logic, allowing
arbitrary roles in number restrictions leads to undecidability, while
decidability for the corresponding variant of \shif is still an open problem.
For convenience, we will refer to \shn with arbitrary roles in number
restrictions as \shnplus.

The undecidability proof uses a reduction of the domino problem~\cite{Ber66}
adapted from~\cite{Baader96b}. This problem asks if, for a set of domino types,
there exists a \emph{tiling} of an $\N^2$ grid such that each point of
the grid is covered with one of the domino types, and adjacent dominoes are
``compatible'' with respect to some predefined criteria.

\begin{Definition}
  A domino system $\mathcal{D} = (D, H, V)$ consists of a non-empty
  set of domino types $D = \{D_1,\ldots,D_n\}$, and of sets of
  horizontally and vertically matching pairs $H \subseteq D \times D$
  and $V \subseteq D \times D$. The problem is to determine if, for a
  given $\mathcal{D}$, there exists a \emph{tiling} of an $\N \times
  \N$ grid such that each point of the grid is covered with a domino
  type in $D$ and all horizontally and vertically adjacent pairs of
  domino types are in $H$ and $V$ respectively, i.e., a mapping $t:\N
  \times \N \rightarrow D$ such that for all $m,n \in \N$,
  $\tuple{t(m,n)}{t(m+1,n)} \in H$ and $\tuple{t(m,n)}{t(m,n+1)} \in
  V$.
\end{Definition}

This problem can be reduced to the
satisfiability of \shnplus-concepts, and the undecidability of the domino
problem implies undecidability of satisfiability of \shnplus-concepts.

Ensuring that a given point satisfies the compatibility conditions is simple for
most logics (using value restrictions and boolean connectives), and applying such
conditions throughout the grid is also simple in a logic such as \shnplus which
can deal with arbitrary axioms.  The crucial difficulty is representing the $\N
\times \N$ grid using ``horizontal'' and ``vertical'' roles $X$ and $Y$, and in
particular forcing the
coincidence of $X \circ Y$ and $Y \circ X$ successors. 
This can be accomplished in \shnplus using an alternating pattern of two
horizontal roles $X_1$ and $X_2$, and two vertical roles $Y_1$ and $Y_2$, with
disjoint primitive concepts $A$, $B$, $C$, and $D$ being used to identify points
in the grid with different combinations of successors. The coincidence of $X
\circ Y$ and $Y \circ X$ successors can then be enforced using number
restrictions on transitive super-roles of each of the four possible combinations
of $X$ and $Y$ roles. A visualisation of the resulting grid and a suitable role
hierarchy is shown in Figure~\ref{fig:grid}, where $S^\oplus_{ij}$ are transitive roles.

\begin{figure}
  \begin{center}
    \parbox[t]{3.2in}{\input{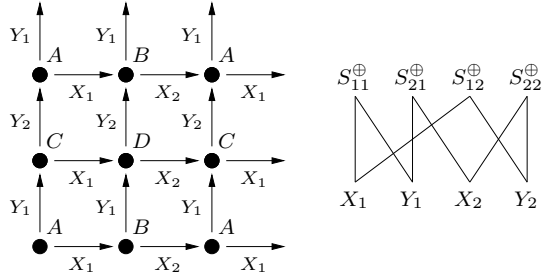}}
    \caption{Visualisation of the grid and role hierarchy.}
    \label{fig:grid}\vspace{-0.5cm}
  \end{center}
\end{figure}

The alternation of $X$ and $Y$ roles in the grid means that one of the
transitive super-roles $S^\oplus_{ij}$ connects each point $(x, y)$ to the points
$(x+1, y)$, $(x, y+1)$ and $(x+1, y+1)$, and to no other points. A number
restriction of the form \atmost{3}{S^\oplus_{ij}} can thus be used to enforce
the necessary coincidence of $X \circ Y$ and $Y \circ X$ successors.
A complete specification of the grid is given by the following axioms:
{\setlength{\arraycolsep}{0em}$$\begin{array}{rcl}
A & \mbox{}\sqsubseteq\mbox{} & \neg B \sqcap \neg C \sqcap \neg
  D \sqcap \some{X_1}{B} \sqcap \some{Y_1}{C}
  \sqcap \atmost{3}{S^\oplus_{11}}\mbox{,} \\
B & \sqsubseteq & \neg A \sqcap \neg C \sqcap \neg
  D \sqcap \some{X_2}{A} \sqcap \some{Y_1}{D}
  \sqcap \atmost{3}{S^\oplus_{21}}\mbox{,} \\
C & \sqsubseteq & \neg A \sqcap \neg B \sqcap \neg
  D \sqcap \some{X_1}{D} \sqcap \some{Y_2}{A}
  \sqcap \atmost{3}{S^\oplus_{12}}\mbox{,} \\
D & \sqsubseteq & \neg A \sqcap \neg B \sqcap \neg
  C \sqcap \some{X_2}{C} \sqcap \some{Y_2}{B}
  \sqcap \atmost{3}{S^\oplus_{22}}\mbox{.}
\end{array}$$}
It only remains to add axioms which encode the local compatibility conditions (as
described in ~\cite{Baader96b}) and to assert that $A$ is subsumed by the
disjunction of all domino types. The \shnplus-concept $A$ is now satisfiable w.r.t.\ the
various axioms (which can be internalised as described in
Lemma~\ref{lemma:terminologies}) iff there is a compatible tiling of the grid.

\newcommand{\DB}{DB\xspace}
\newcommand{\Textand}{\ensuremath{\mathbin{\mathsf{and}}}}
\newcommand{\Textor}{\ensuremath{\mathbin{\mathsf{or}}}}
\newcommand{\Textimplies}{\ensuremath{\mathbin{\mathsf{implies}}}}
\newcommand{\Iff}{\ensuremath{\mathbin{\mathsf{iff}}}}
\newcommand{\Bd}{\ensuremath{b}}
\newcommand{\Key}[1]{\texttt{#1}}

\section{Implementation and Optimisation}
\label{09-sec:introduction}

The development of the \si family of DLs has been motivated by the
desire to implement systems with good typical case performance. As
discussed in Section~\ref{sec:blocking}, this
is achieved in part through the design of the logics and algorithms
themselves, in particular by using transitive roles and by reasoning
with number restrictions directly, rather
than via encodings.
Another important feature of these algorithms is that their relative
simplicity facilitates the application of a range of optimisation
techniques. Several systems based on \s logics have now been
implemented (e.g., \Fact~\cite{Horrocks98d},
\Dlp~\cite{Patel-Schneider98b} and RACE~\cite{Haarslev99a}), and have demonstrated that suitable
optimisation techniques can lead to a dramatic improvement in the
performance of the algorithms when used in realistic applications. A
system based on the \shif logic has also been implemented
(\iFaCT~\cite{Horrocks99c}) and has been shown to be similarly
amenable to optimisation.

\DL systems are typically used to classify a \KB, and the optimisation
techniques used in such systems can be divided into
four categories based on the stage of
the classification process at which they are applied.
\begin{enumerate}
\item Preprocessing optimisations that try to modify the \KB so that
  classification and subsumption testing are easier.
\item Partial ordering optimisations that try to minimise the number
  of subsumption tests required in order to classify the \KB.
\item Subsumption optimisations that try to avoid performing
  a potentially expensive satisfiability test, usually by substituting
  a cheaper test.
\item Satisfiability optimisations that try to improve the typical
  case performance of the underlying satisfiability testing algorithm.
\end{enumerate}
Many optimisations in the first three categories are relatively
independent of the underlying subsumption (satisfiability) testing
algorithm and could be applied to any \DL system. As we are mostly
concerned with algorithms for the \si family of \DL{}s we will
concentrate on the fourth kind of optimisation, those that try to
improve the performance of the algorithm itself. Most of these are
aimed at reducing the size of the search space explored by the
algorithm as a result of applying non-deterministic tableaux expansion
rules.

\subsection{ Semantic branching search}
\label{09-sec:sembranch}

Implementations of the algorithms described in the previous sections
typically use a search technique called
\emph{syntactic branching}. When expanding the label of a node $x$,
syntactic branching works by choosing an unexpanded disjunction $(C_1
\sqcup \ldots \sqcup C_n)$ in $\Lab(x)$ and searching the different
models obtained by adding each of the disjuncts $C_1$, \ldots, $C_n$
to $\Lab(x)$~\cite{GiSe96}.  As the alternative branches of the search
tree are not disjoint, there is nothing to prevent the recurrence of
an unsatisfiable disjunct in different branches. The resulting wasted
expansion could be costly if discovering the unsatisfiability requires
the solution of a complex sub-problem. For example, tableaux expansion
of a node $x$, where $\{(A \sqcup B),(A \sqcup C)\} \subseteq \Lab(x)$
and $A$ is an unsatisfiable concept, could lead to the search pattern
shown in Figure~\ref{09-fig:synbranch}, in which the unsatisfiability
of $\Lab(x) \cup \{A\}$ must be demonstrated twice.

\begin{figure}
\centering
\parbox{3.5in}{\hspace*{-1.15in}\input{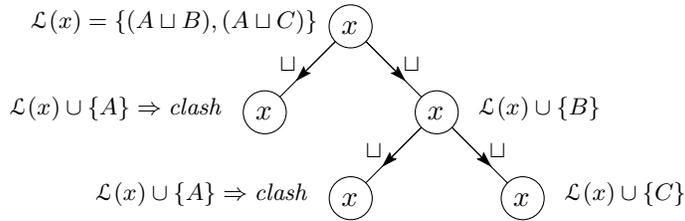}}
\caption{Syntactic branching search}
\label{09-fig:synbranch}
\end{figure}

This problem can be dealt with by using a \emph{semantic branching}
technique adapted from the Davis-Putnam-Logemann-Loveland procedure
(DPL) commonly used to solve propositional satisfiability (SAT)
problems~\cite{DaLL62,Free96}. Instead of choosing an unexpanded
disjunction in $\Lab(x)$, a single disjunct $D$ is chosen from one of
the unexpanded disjunctions in $\Lab(x)$. The two possible sub-trees
obtained by adding either $D$ or $\Not D$ to $\Lab(x)$ are then
searched. Because the two sub-trees are strictly disjoint, there is no
possibility of wasted search as in syntactic branching. Note that the
order in which the two branches are explored is irrelevant from a
theoretical viewpoint, but may offer further optimisation
possibilities (see Section~\ref{09-sec:heuristics}).

Semantic branching search has the additional advantage that a great
deal is known about the implementation and optimisation of the DPL
algorithm. In particular, both \emph{local simplification} (see
Section~\ref{09-sec:bcp}) and \emph{heuristic guided search} (see
Section~\ref{09-sec:heuristics}) can be used to try to minimise the
size of the search tree (although it should be noted that both these
techniques can also be adapted for use with syntactic branching
search).

There are also some disadvantages to semantic branching search.
Firstly, it is possible that performance could be degraded by adding
the negated disjunct in the second branch of the search tree, for
example if the disjunct is a very large or complex concept.  However
this does not seem to be a serious problem in practice, with semantic
branching rarely exhibiting significantly worse performance than
syntactic branching. Secondly, its effectiveness is problem
dependent. It is most effective with randomly generated problems,
particularly those that are over-constrained (likely to be
unsatisfiable)~\cite{HoPa99}. It is also effective with some of the
hand crafted problems from the Tableaux'98 benchmark
suite~\cite{HeSc96,BaHe98}. However it is of little benefit when
classifying realistic \KB{}s~\cite{HoPa98b}.

\subsection{ Local simplification}
\label{09-sec:bcp}

Local simplification is another technique used to reduce the size of
the search space resulting from the application of non-deterministic
expansion rules. Before any non-deterministic expansion of a node
label $\Lab(x)$ is performed, disjunctions in $\Lab(x)$ are examined,
and if possible simplified. The simplification most commonly used is
to deterministically expand disjunctions in $\Lab(x)$ that present
only one expansion possibility and to detect a clash when a
disjunction in $\Lab(x)$ has no expansion possibilities.
This simplification has been called \emph{boolean constraint propagation}
(BCP)~\cite{Free95}.  In effect, the inference rule
$$\frac{\Not C_1, \ldots, \Not C_n, C_1 \sqcup \ldots \sqcup C_n
  \sqcup D}{D}$$
is being used to simplify the conjunctive concept represented by $\Lab(x)$.
For example, given a node $x$ such that
$$\{(C \sqcup (D_1 \sqcap D_2)),(\Not D_1 \sqcup \Not D_2 \sqcup
C),\Not C\} \subseteq \Lab(x),$$
BCP deterministically expands the
disjunction $(C \sqcup (D_1 \sqcap D_2))$, adding $(D_1 \sqcap D_2)$
to $\Lab(x)$, because $\Not C \in \Lab(x)$. The deterministic expansion
of $(D_1 \sqcap D_2)$ adds both $D_1$ and $D_2$ to $\Lab(x)$, allowing
BCP to identify $(\Not D_1 \sqcup \Not D_2 \sqcup C)$ as a clash (without any
branching having occurred), because $\{D_1, D_2, \Not C\}
\subseteq \Lab(x)$.

BCP simplification is usually described as an integral part of SAT
based algorithms~\cite{GiSe96}, but it can also be used with syntactic
branching. However, it is more effective with semantic branching as
the negated concepts introduced by failed branches can result in
additional simplifications. Taking the above example of $\{(A \sqcup
B),(A \sqcup C)\} \subseteq \Lab(x)$, adding $\Not A$ to $\Lab(x)$
allows BCP to deterministically expand both of the disjunctions using
the simplifications $(A \sqcup B) \Textand \Not A
\rightarrow B$ and $(A \sqcup C)\Textand \Not A
\rightarrow C$. The reduced search space resulting from the
combination of semantic branching and BCP is shown in
Figure~\ref{09-fig:sembranch}.

\begin{figure}
\centering
\parbox{3.45in}{\hspace*{-1.15in}\input{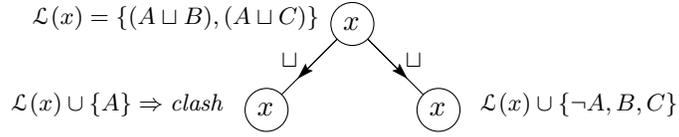}}
\caption{Semantic branching search}
\label{09-fig:sembranch}
\end{figure}

Local simplification has the advantage that it can never increase the
size of the search space and can thus only degrade performance to the
extent of the overhead required to perform the
simplification. Minimising this overhead does, however, require
complex data structures~\cite{Free95}, particularly in a
modal/description logic setting.

As with semantic branching, effectiveness is problem dependent, the
optimisation being most effective with over-constrained randomly
generated problems~\cite{HoPa98b}.

\subsection{ Dependency directed backtracking}
\label{09-sec:backjumping}

Inherent unsatisfiability concealed in sub-problems can lead to large
amounts of unproductive backtracking search, sometimes called
thrashing. For example, expanding a node $x$ (using semantic
branching), where
$$\Lab(x) = \{(C_1 \sqcup D_1), \ldots, (C_n \sqcup D_n), \Some{R}{(A
  \sqcap B)}, \All{R}{\Not A}\},$$
could lead to the fruitless exploration of $2^n$ possible
$R$-successors of $x$ before the inherent unsatisfiability is
discovered.  The search
tree resulting from the tableaux expansion is illustrated in
Figure~\ref{09-fig:backjumping}.

\begin{figure}
\centering
\parbox{4.85in}{\hspace*{-0.4in}\input{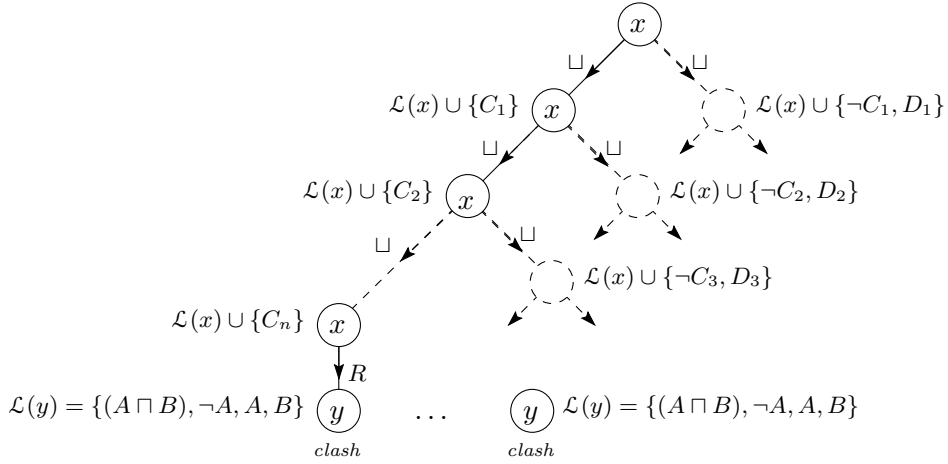}}
\caption{Thrashing in backtracking search}
\label{09-fig:backjumping}
\end{figure}

This problem can be addressed by adapting a form of dependency
directed backtracking called \emph{backjumping}, which has been used
in solving constraint satisfiability problems~\cite{Bake95} (a
similar technique was also used in the HARP theorem
prover~\cite{OpSu88}).  Backjumping works by labelling each
concept in a node label with a dependency set indicating the branching
points on which it depends. A concept $C \in \Lab(x)$ depends on a
branching point if $C$ was added to $\Lab(x)$ at the branching point
or if $C \in \Lab(x)$ was generated by an expansion rule (including
simplification) that depends on another concept $D \in \Lab(y)$, and
$D \in \Lab(y)$ depends on the branching point.  A concept $C \in
\Lab(x)$ depends on a concept $D \in \Lab(y)$ when $C$ was added to
$\Lab(x)$ by a deterministic expansion that used $D \in \Lab(y)$.  For
example, if $A \in \Lab(x)$ was derived from the expansion of $(A
\sqcap B) \in \Lab(x)$, then $A \in \Lab(x)$ depends on $(A \sqcap B)
\in \Lab(x)$. 

When a clash is discovered, the dependency sets of the clashing
concepts can be used to identify the most recent branching point where
exploring the other branch might alleviate the cause of the clash. It
is then possible to jump back over intervening branching points
\emph{without} exploring any alternative branches. 
Let us consider the earlier example and suppose that $\Some{R}{(A
  \sqcap B)}$ has a dependency set $\mathbf{D}_i$ and $\All{R}{\Not A}$
has a dependency set $\mathbf{D}_j$. The search proceeds until $C_1
\ldots C_n$ have been added to $\Lab(x)$, when $\Some{R}{(A \sqcap B)}$ and $\All{R}{\Not A}$ are deterministically
expanded and a clash occurs in $\Lab(y)$ between the $A$ derived from
$\Some{R}{(A \sqcap B)}$ and the $\Not A$ derived from $\All{R}{\Not
  A}$. As these derivations were both deterministic, the dependency
sets will be $\mathbf{D}_i$ and $\mathbf{D}_j$ respectively, and so
$\mathbf{D}_i \cup \mathbf{D}_j$ is returned. This set cannot include
the branching points where $C_1
\ldots C_n$ were added to $\Lab(x)$ as $\mathbf{D}_i$ and
$\mathbf{D}_j$ were defined before these branching points were
reached. The algorithm can therefore backtrack through each of the preceding $n$ branching points without
exploring the second branches, and will continue
to backtrack until it reaches the branching point equal to the maximum
value in $\mathbf{D}_i \cup \mathbf{D}_j$ (if $\mathbf{D}_i =
\mathbf{D}_j = \emptyset$, then the algorithm will backtrack through
all branching points and return ``unsatisfiable'').
Figure~\ref{09-fig:pruning} illustrates the pruned search tree, with
the number of $R$-successors explored being reduced by an exponential number.

\begin{figure}
\centering
\parbox{3.85in}{\hspace*{-0.85in}\input{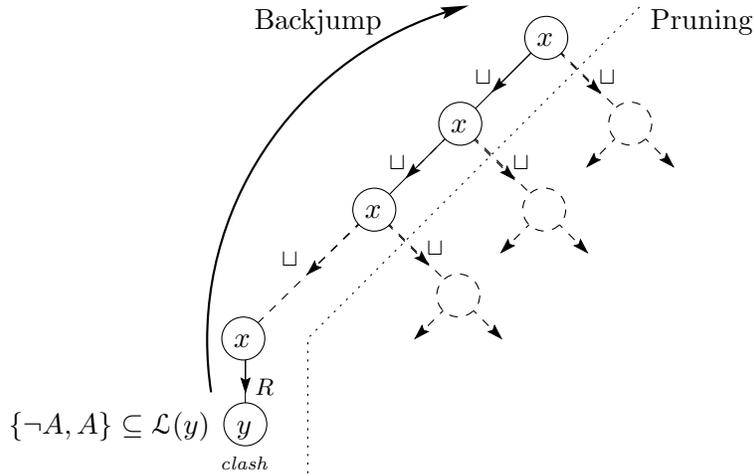}}
\caption{Pruning the search using backjumping}
\label{09-fig:pruning}
\end{figure}

Backjumping can also be used with syntactic branching, but the
procedure is slightly more complex as there may be more than two
possible choices at a given branching point, and the dependency set of
the disjunction being expanded must also be taken into account.

Like local simplification, backjumping can never increase the size of
the search space. Moreover, it can lead to a dramatic reduction in the size of the search
  tree and thus a huge performance improvement. For example, when
   using either \Fact or \Dlp with
  backjumping disabled in order to classify a large ($\approx$3,000 concept) \KB derived from 
  the European \Galen project~\cite{Rector93a},
 single satisfiability tests were encountered
  that could not be solved even after several weeks of CPU
  time. Classifying the same \KB with backjumping enabled takes less
  than 100s of CPU
  time for either \Fact or \Dlp~\cite{HoPa99}.

Backjumping's only disadvantage is the overhead of propagating and storing the
dependency sets. This can be alleviated to some extent
by using a pointer based implementation so that propagating a
dependency set only requires the copying of a pointer.

\subsection{ Heuristic guided search}
\label{09-sec:heuristics}

Heuristic techniques can be used to guide the search in a way that
tries to minimise the size of the search tree. A method that is widely
used in DPL SAT algorithms is to branch on the disjunct that has the
\emph{Maximum number of Occurrences in disjunctions of Minimum Size}---the
well known MOMS heuristic~\cite{Free95}. By choosing a disjunct
that occurs frequently in small disjunctions, the MOMS heuristic tries
to maximise the effect of BCP. For example, if the label of a node $x$
contains the unexpanded disjunctions $C \sqcup D_1, \ldots, C \sqcup
D_n$, then branching on $C$ leads to their deterministic expansion in
a single step: when $C$ is added to $\Lab(x)$, all of the disjunctions
are fully expanded and when $\Not C$ is added to $\Lab(x)$, BCP will
expand all of the disjunctions, causing $D_1, \ldots, D_n$ to be added
to $\Lab(x)$. Branching first on any of $D_1, \ldots, D_n$, on the
other hand, would only cause a single disjunction to be expanded.

The MOMS value for a candidate concept $C$ is computed simply by
counting the number of times $C$ or its negation occur in minimally
sized disjunctions. There are several variants of this heuristic,
including the heuristic from Jeroslow and Wang~\cite{JeWa90}.  The
Jeroslow and Wang heuristic considers all occurrences of a disjunct,
weighting them according to the size of the disjunction in which they
occur. The heuristic then selects the disjunct with the highest
overall weighting, again with the objective of maximising the effect
of BCP and reducing the size of the search tree.

When a disjunct $C$ has been selected from the disjunctions in
$\Lab(x)$, a BCP maximising heuristic can also be used to determine
the order in which the two possible branches, $\Lab(x) \cup \{C\}$ and
$\Lab(x) \cup \{\Not C\}$, are explored. This is done by separating
the two components of the heuristic weighting contributed by
occurrences of $C$ and $\Not C$, trying $\Lab(x) \cup \{C\}$ first if
$C$ made the \emph{smallest} contribution, and trying $\Lab(x) \cup
\{\Not C\}$ first otherwise. The intention is to prune the search tree
by maximising BCP in the first branch.

Unfortunately MOMS-style heuristics can interact adversely with the
backjumping optimisation because they do not take dependency
information into account. This was first discovered in the \Fact
system, when it was noticed that using MOMS heuristic often led to
much worse performance. The cause of this phenomenon turned out to be
the fact that, without the heuristic, the data structures used in the
implementation naturally led to ``older'' disjunctions (those
dependent on earlier branching points) being expanded before ``newer''
ones, and this led to more effective pruning if a clash was
discovered. Using the heuristic disturbed this ordering and reduced
the effectiveness of backjumping~\cite{Horr97b}.

Moreover, MOMS-style heuristics are of little value themselves in
description logic systems because they rely for their effectiveness on
finding the same disjuncts recurring in multiple unexpanded
disjunctions: this is likely in hard propositional problems, where the
disjuncts are propositional variables, and where the number of
different variables is usually small compared to the number of
disjunctive clauses (otherwise problems would, in general, be
trivially satisfiable); it is unlikely in concept satisfiability
problems, where the disjuncts are (possibly non-atomic) concepts, and
where the number of different concepts is usually large compared to
the number of disjunctive clauses. As a result, these heuristics will
often discover that all disjuncts have similar or equal priorities,
and the guidance they provide is not particularly useful.

An alternative strategy is to employ an {\em oldest-first} heuristic
that tries to maximise the effectiveness of backjumping by using
dependency sets to guide the expansion~\cite{HoPa99}. When choosing a
disjunct on which to branch, the heuristic first selects those
disjunctions that depend on the least recent branching points (i.e.,
those with minimal maximum values in their dependency sets), and then
selects a disjunct from one of these disjunctions.  This can be
combined with the use of a BCP maximising heuristic, such as the
Jeroslow and Wang heuristic, to select the disjunct from amongst the
selected disjunctions.

The oldest-first heuristic can also be used to advantage when
selecting the order in which existential role restrictions, and the
labels of the $R$-successors which they generate, are expanded. One
possible technique is to use the heuristic to select an unexpanded
existential role restriction $\Some{R}{C}$ from the label of a node
$x$, apply the $\exists$-rule and the $\forall$-rule as necessary, and
expand the label of the resulting $R$-successor. If the expansion results
in a clash, then the algorithm will backtrack; if it does not, then
continue selecting and expanding existential role restrictions from
$\Lab(x)$ until it is fully expanded. A better technique is to first
apply the $\exists$-rule and the $\forall$-rule exhaustively, creating
a set of successor nodes. The order in which to expand these successors
can then be based on the minimal maximum values in the dependency sets
of all the concepts in their label, some of which may be due to
universal role restrictions in $\Lab(x)$.

The main advantage of heuristics is that they can be used to
complement other optimisations. The MOMS and Jeroslow and Wang
heuristics, for example, are designed to increase the effectiveness of
BCP while the oldest-first heuristic is designed to increase the
effectiveness of backjumping. They can also be selected and tuned to
take advantage of the kinds of problem that are to be solved (if this
is known). The BCP maximisation heuristics, for example, are generally
quite effective with large randomly generated and hand crafted
problems, whereas the oldest-first heuristic is more effective when
classifying realistic \KB{}s.

Unfortunately heuristics also have several disadvantages. They can add
a significant overhead as the heuristic function may be expensive to
evaluate and may need to be reevaluated at each branching
point. Moreover, they may not improve performance, and may
significantly degrade it, for example by interacting adversely with
other optimisations, by increasing the frequency with which pathological
worst cases can be expected to occur in generally easy problem sets.

\subsection{ Caching satisfiability status}
\label{09-sec:caching}

During a satisfiability check there may be many successor nodes
created.  Some of these nodes can be very similar, particularly as the
labels of the $R$-successors for a node $x$ each contain the same
concepts derived from the universal role restrictions in
$\Lab(x)$. Systems such as \Dlp take advantage of this similarity by
caching the satisfiability status of the sets of concepts with which
node labels are initialised when they are created. The tableaux
expansion of a node can then be avoided if the satisfiability status
of its initial set of concepts is found in the cache.

However, this technique depends on the logic having the property that
the satisfiability of a node is completely determined by its initial
label set, and, due to the possible presence of inverse roles, \si
logics do not have this property. For example, if the expansion of a
node $x$ generates an $R$-successor node $y$, with
$\Lab(y)=\{\all{R^-}{C}\}$, then the satisfiability of $y$ clearly
also depends on the set of concepts in $\Lab(x)$. Similar problems could
arise in the case where $\Lab(y)$ contains number restriction
concepts.

If it is possible to solve these problems, then caching may be a very
effective technique for \si logics, as it has been shown to be in the
\Dlp system with a logic that does not support inverse roles. Caching
is particularly useful in \KB classification as cached values can be retained across
multiple satisfiability tests. It can also be effective
with both satisfiable and unsatisfiable problems, unlike many other
optimisation techniques that are primarily aimed at speeding up the
detection of unsatisfiability.

The main disadvantage with caching is the storage overhead incurred by
retaining node labels (and perhaps additional information in the case
of \si logics) and their satisfiability status throughout a
satisfiability test (or longer, if the results are to be used in later
satisfiability tests). An additional problem is that it interacts
adversely with the backjumping optimisation as the dependency
information required for backjumping cannot be effectively calculated
for nodes that are found to be unsatisfiable as a result of a cache
lookup. Although the set of concepts in the initial label of such a
node is the same as that of the expanded node whose (un)satisfiability
status has been cached, the dependency sets attached to the concepts
that made up the two labels may not be the same. However, a weaker
form of backjumping can still be performed by taking the dependency
set of the unsatisfiable node to be the union of the dependency sets
from the concepts in its label.

\section{Discussion}

A new DL system is being implemented based on the \shiq algorithm we have
developed from the \shif-algorithm described in
Section~\ref{sec:shin-algo}~\cite{Horrocks99j}.  Pending the completion of this project, the
existing \Fact system~\cite{Horrocks98c} has been modified to deal with inverse
roles using the \shif blocking strategy, the resulting system being
referred to as \IFact.

\IFact has been used to conduct some initial experiments with a terminology
representing (fragments of) database schemata and inter schema assertions from a
data warehousing application~\cite{Calvanese98c} 
(a slightly simplified version of the proposed encoding was used to generate 
\shif terminologies).
\IFact is able to classify this terminology,
which contains 19 concepts and 42 axioms, in less than 0.1s of (266MHz Pentium)
CPU time. In contrast, eliminating inverse roles using an embedding
technique~\cite{CaGiaRo98} gives an equisatisfiable \Fact terminology with an additional
84 axioms, but one which \Fact is unable to classify in 12 hours of CPU time.
As discussed in Section~\ref{sec:blocking}, an extension of the embedding technique can be used to eliminate number
restrictions~\cite{DeGiacomo95b}, but requires a target logic which
supports the transitive \emph{closure} of roles, i.e., \CPDL. The even larger
number of axioms that this embedding would introduce makes it unlikely that
tractable reasoning could be performed on the resulting terminology. Moreover,
we are not aware of any algorithm for \CPDL which does not employ a so-called
\emph{look behind analytical cut}~\cite{DeGiaMass98}, the application of which introduces
considerable additional non-determinism. It seems inevitable that this would
lead to a further degradation in empirical tractability.

The \DL \shiq will allow the above mentioned encoding of database
schemata to be fully captured using qualified number restrictions.
Future work will include completing the implementation of the \shiq
algorithm, testing its behaviour in this kind of application and
investigating new techniques for improving its empirical tractability.

\end{document}
